\documentstyle[aps,twocolumn,floats]{revtex}
\begin{document}

\tighten
\draft
\twocolumn[\hsize\textwidth\columnwidth\hsize\csname 
@twocolumnfalse\endcsname

\title{Back Reaction And Local Cosmological Expansion Rate}

\author{Ghazal Geshnizjani$^{1)}$ and Robert Brandenberger$^{2)}$}

\address{1) Department of Physics, Brown University, Providence, RI 02912, 
USA\\
and\\
Department of Astrophysical Sciences, Princeton University,\\ 
Princeton, NJ 08544, USA\\
E-mail: ghazal@het.brown.edu}

\address{2) TH Division, CERN, CH-1211 Geneva 23, Switzerland,\\
Institut d'Astrophysique de Paris, 98bis Blvd. Arago, F-75014 Paris, France\\
and
Department of Physics, Brown University, Providence, RI 02912, USA\\
E-mail: rhb@het.brown.edu}

\maketitle

\begin{abstract}

We calculate the back reaction of cosmological perturbations on
a general relativistic variable which measures the local expansion
rate of the Universe. Specifically, we consider a cosmological
model in which matter is described by a single field. 
We analyze back reaction both in a matter dominated Universe and in
a phase of scalar field-driven chaotic inflation. In both cases,
we find that the leading infrared terms contributing to the
back reaction vanish when the local expansion rate is measured
at a fixed value of the matter field which is used as a clock,
whereas they do not appear to vanish if the expansion rate is
evaluated at a fixed value of the background time. We discuss
possible implications for more realistic models with a more complicated matter
sector.  

\end{abstract}

\vspace*{1cm}
]

\section{Introduction}

Because of the nonlinear nature of the Einstein equations, linear
fluctuations about a homogeneous and isotropic FRW cosmology will
back-react on the background on which they live. 
The back-reaction of short wavelength gravitational
waves was studied a long time ago by Brill, Hartle and Isaacson
 \cite{BHI}. In this approach, the
back-reaction of fluctuations on the spatially averaged metric 
can be described by an effective energy-momentum
tensor which contains the spatially averaged terms in the Einstein
tensor which are second order in the amplitude of the
fluctuations.  More recently, Tsamis and Woodard \cite{WT} have
initiated a detailed study of the back-reaction of long wavelength
gravitational waves in a de Sitter background. Scalar-type metric
fluctuations (called ``cosmological fluctuations'' in what follows) 
are also expected to back-react on the
background space-time. Since in the context of inflationary and
post-inflationary cosmology the scalar metric fluctuations are
believed to dominate over the effects of gravitational waves, the
back-reaction of these cosmological perturbations are
expected to give the dominating back-reaction effect. 

In \cite{bran2,bran3}, the formalism of \cite{BHI} was generalized
to describe the back-reaction of cosmological perturbations on
the spatially averaged metric. On this basis, it was
argued \cite{bran2,bran3,aw0} that gravitational back-reaction in 
scalar field-driven inflationary models, calculated up to quadratic order 
in perturbations and to leading
order in the long wavelength expansion and in the slow roll approximation could
decrease the expansion rate of the universe
and potentially solve the cosmological constant problem \cite{bran4} (see
also \cite{atw} for similar results in the context of back-reaction
studies of gravitational waves). The formalism of \cite{bran2,bran3} is
covariant under first order space-time diffeomorphisms 
\footnote{The back-reaction of small-scale (i.e. smaller than
the Hubble radius) cosmological perturbations has been
considered (without taking into account the issues of gauge freedom) in
\cite{Futamase1,Futamase2,BF,Futamase3,SH,Ruus,Jelle,Buchert}. 
This problem has also 
been considered in the context of Newtonian cosmological 
perturbation theory in \cite{BE,TF}. More recently,
Nambu \cite{Nambu1,Nambu2,Nambu3}
has initiated a program to compute back-reaction effects on the
spatially averaged metric using the renormalization group method.}.

However, as emphasized by Unruh \cite{unruh} \footnote{We are also grateful
to Andrei Linde and Alan Guth for detailed private discussions on
these points.}, the approach of
\cite{bran2,bran3,WT} is deficient in several respects. First of all,
due to the nonlinear nature of the Einstein equations, calculating
an ``observable'' from the spatially averaged metric will not in
general give the same result as calculating the spatially averaged value
of the observable. More importantly, the spatially averaged metric is
not a local physical observable. Thus, to take into account
the deficiencies of the previous work on gravitational back-reaction,
we must identify a local physical variable which describes the
expansion rate of the Universe, calculate the back-reaction
of cosmological perturbations on this quantity, and then
take the spatially averaged value of the result. It is important
to fix the hypersurface of averaging by a clear physical
prescription in order to remove the possibility of being
misled by effects which are second order gauge artifacts.

In this paper, we propose an implementation of this approach.
We focus on the variable which yields the general relativistic
definition of the local expansion rate, calculate this quantity to second order in the amplitude of
the cosmological fluctuations, in terms of a time variable defined by an unambiguous physical prescription.
For simplicity, we assume that matter is described by a single
field (either a hydrodynamical field or a single scalar field).

We study two examples, first a matter dominated Universe, and
second the inflationary phase of a cosmology dominated by
a single scalar field. In both cases we find that the leading
infrared contributions to the back-reaction on the local
expansion rate of the Universe vanish, in contrast to
the findings of the initial work on gravitational back-reaction
of cosmological fluctuations \cite{bran2,bran3,aw0}, and
confirming the analysis of \cite{unruh}.
We thus confirm the conclusions reached recently in 
\cite{aw2,aw3} where a different variable related to the local
expansion rate is proposed, and different techniques to
evaluate this variable are used. 

Note that when evaluated at a fixed background time, our leading
infrared back-reaction terms give a non-vanishing contribution. This
leads us to the conjecture that in more realistic
models in which a second field
is present to determine time (e.g. the microwave background), the
leading infrared back-reaction terms will not vanish.

\section{A local observable}

For a general perfect fluid flow in a curved
space-time we consider the velocity four-vector field $u^{\alpha}$
tangential to a family of world lines. In the
context of cosmology, we can always define a
preferred family of world lines representing the motion of a set of
comoving observers. In the case of hydrodynamical matter, this is
easy since the energy-momentum tensor is already defined in terms of
a velocity four-vector field. Also in the case of scalar field matter,
a corresponding four-vector field can be defined, although a
bit more care is required to obtain a consistent
definition. In both cases we have
\begin{equation} \label{norm}
       u^{\alpha}u_{\alpha} \, = \, 1 \, ,
\end{equation}
where $\alpha$ runs over the space-time indices. The projection
tensor onto tangential three-surfaces orthogonal to $u^{\alpha}$ is
\begin{equation}
h_{\alpha\beta} \, = \, g_{\alpha\beta} - u_{\alpha}u_{\beta} \, .
\end{equation}

The first covariant derivative  of
the four-velocity can be decomposed as (see e.g. \cite{EB} for details )
\begin{equation}
u_{\alpha;\beta} \, = \, \omega_{\alpha\beta} + \sigma_{\alpha\beta} 
+ \frac{1}{3}\Theta h_{\alpha\beta}-\dot{u}_{\alpha}u_{\beta} \, .
\end{equation}
Here 
\begin{equation}
\Theta \, \equiv \, u^{\alpha}_{\,;\alpha}
\end{equation}
is the local expansion rate of the tangential surfaces
orthogonal to the fluid flow, $\omega$ is the vorticity tensor
(with $\omega_{\alpha\beta}u^{\beta} = 0$),  
and $\sigma$ is the shear tensor (satisfying 
$\sigma_{\alpha\beta}u^{\beta}=0, \, \sigma^{\alpha}_{\alpha}=0$).

For a homogeneous Universe
with scale factor $a(t)$ the Hubble expansion rate $H$ is
\begin{equation} \label{Hubble}
H \, \equiv \, \dot{a}/a \, = \, \frac{1}{3}\Theta \, .
\end{equation}
For a cosmological model with fluctuations, $\Theta$ is local in space
and time. We will use $\Theta$ to define the local expansion rate $\dot{S}/S$ 
via the local analog of (\ref{Hubble}), namely via the equation
\begin{equation} \label{localHubble}
\dot{S}/S \, \equiv \,\frac{1}{3}\Theta \, .
\end{equation}
The quantity $\dot{S}/S$ is a much better measure of the locally
measured expansion rate in a Universe with fluctuations than the
Hubble expansion rate used in \cite{bran2,bran3} determined via
the spatially averaged metric including back-reaction. It is
a mathematically simpler object than the variable recently
introduced in \cite{aw2} which involves the integral 
along the past light cone of the observation point.  
If we are interested in evaluating the expansion rate for a typical
observer, we propose to take the spatial average of the local
expansion rate defined via (\ref{localHubble}).

Now that we have defined the observable we are
interested in, the procedure will be as follows.
First, we must determine the velocity four-vector field
$u^{\alpha}$ for the models we are interested in. Then,
we use the Einstein equations to express $u^{\alpha}$
in terms of the metric perturbation. Taking the 
relative amplitude of the metric fluctuations as the expansion
parameter, we then calculate
$\Theta$, our local measure of the Hubble expansion rate, to
second order. After evaluating the result on a physically determined hypersurface
we can then study the back-reaction of cosmological fluctuations
on the locally measured Hubble expansion rate. In this paper
we will focus on the leading infrared contributions to back-reaction,
the terms found to dominate the back-reaction effects in
\cite{bran2,bran3,aw0}.

\section{Deriving the Expansion Rate for Scalar Metric Perturbations}

In this section we consider a model with hydrodynamical matter. 
Starting from the expression for the metric to linear order
in the fluctuations $\Phi$ (see \cite{RA} for a detailed review), we
determine the velocity four vector field $u^{\alpha}$ to the
order required to analyze the leading infrared terms
in the back-reaction to quadratic order. To obtain the full
back-reaction terms (including terms which dominate in the
ultraviolet but are negligible in the infrared) we should
calculate  $u^{\alpha}$ consistently up to second order.
However, if we are only interested in the leading infrared terms, it
is sufficient to keep all the terms quadratic in $\Phi$ but not
containing any spatial gradients. 

In order to obtain the complete result for gravitational
back-reaction we would
have to look at the Einstein equations for a perfect fluid
with energy density $\rho$ and pressure $P$,
\begin{equation}
G_{\mu\nu} =(P+\rho)u_{\mu}u_{\nu}-Pg_{\mu\nu}
\end{equation}
(in units in which $8 \pi G = 1$), which, since $G^{\mu}_{\mu}=R$, 
will yield
\begin{eqnarray}
R&=&\rho-3P \\
\rho&=&u^{\mu}G_{\mu\nu}u^{\nu} \, ,
\end{eqnarray}
and lead to an equation that can be solved perturbatively
to any desired order for $u_{i}$:
\begin{equation} \label{einstein}
G^{0}_{i}=\frac{4}{3}u^{\mu}G_{\mu\nu}u^{\nu}u^{0}u_{i}+Ru^{0}u_{i}
\end{equation}
However, as mentioned above, here we just use the results for $u^{i}$
which are of linear order. 
Since we will calculate the divergence of $u^{\mu}$, our
prescription implies that we are
ignoring some of the extra second order gradient terms .
The result can also be used for scalar fields if we define the $u$
vector field in a proper way. 

For an unperturbed Robertson-Walker
metric, the four-velocity field $u$ in comoving coordinates would be
\begin{equation} 
u^{\mu}=(1,0,0,0)
\end{equation}
In linear perturbation theory, and in the
case of simple forms of matter (such as a single fluid or a single scalar
field) for which there is to linear order no anisotropic
stress, the metric (in longitudinal gauge) can be
written as
\begin{eqnarray} \label{metric}
ds^{2}&=& a(\eta)^2\bigl((1+2\Phi)d\eta^{2} - 
(1-2\Psi)\gamma _{ij}dx^{i}dx^{j}\bigr),\\
\gamma_{ij}&=&\delta_{ij}[1+\frac{1}{4}{\cal
K}(x^{2}+y^{2}+z^{2})]
\end{eqnarray}
where ${\cal K}=0,1,-1$  depending on whether the three-dimensional
space corresponding to the hypersurface $t = {\rm const.}$ is flat,
closed or open. In this paper we will take it to be zero in order to
simplify the calculations. The time variable $\eta$ appearing in
(\ref{metric}) is conformal time and is related to the physical
time $t$ via $d\eta=a^{-1}dt$.
For the forms of matter
considered here, $\Psi=\Phi$ at linear order \footnote{Even if we did not 
make the assumption $\Psi = \Phi$, it turns out that at second order 
all infrared terms
depending on $\Psi - \Phi$ will drop out.}.  As discussed e.g.
in \cite{RA}, in longitudinal gauge the spatial components of the 
four-velocity vector field are related to $\Phi$ via
\begin{equation}
\delta u_{i} \, = \, -a^{-2}({\cal H} ^{2}- {\cal H} ^{\prime}+ {\cal
K})^{-1}(a\Phi)^{\prime}_{,i}
\end{equation}
where a prime denotes
differentiation with respect to $\eta$ and ${\cal H}=a^{\prime}/a$.
Using equation (\ref{norm}) we can derive the expression for the time 
component of $u^{\alpha}$ in terms of $\Phi$:
\begin{equation}
u^{0}(\eta)=a^{-1}(1-\Phi+\frac{3}{2}\Phi^{2}) 
\end{equation} 

Now that we have all components of $u^{\alpha}$, we take the covariant
derivative of it and retain all $\Phi$ dependence up to second
order \footnote{This is in the philosophy of the general back-reaction
approach in which it is assumed that the fluctuations of the metric
and matter satisfy the linear perturbation equations, and we compute
their back-reaction on physical quantities to second order. It is not
a consistent second order perturbative formalism.} These second order
terms principally come from the Christoffel symbols. Other second
order terms (which as mentioned before are gradient terms) could be
added to the ones computed here if we were to solve the Einstein equations 
in the form (\ref{einstein}) beyond linear
order. A straightforward calculation yields
\begin{eqnarray} \label{expform}
\Theta \, &=& \, 3\frac{a^{\prime}}{a^{2}}(1-\Phi+\frac{3}{2}\Phi^{2})
-3\frac{\Phi^{\prime}}{a} \nonumber \\
&+& \, \frac{(a\partial_{i}\Phi)^{\prime}
(\partial_{i}\Phi)+(a\partial_{i}^{2}\Phi)^{\prime}}{a^{2}({\cal
H}^{\prime}-{\cal H}^{2})} \, .
\end{eqnarray}
Since $\dot{a}/a=a^{\prime}/a^{2}$, we can immediately read off the extra
terms contributing to the 
local expansion rate which result from the presence of
cosmological fluctuations. Upon spatial averaging at a fixed conformal
time, the terms linear in $\Phi$ drop out. Hence, it follows
that if evaluated at a fixed conformal time, infrared modes on average 
lead to an increase in the expansion rate compared to what would be
obtained at the same conformal time in the absence of metric
fluctuations. Whether this is a physically measurable effect from an 
observational point of view will be discussed in more depth in  
following sections.
 
\section{Expansion Rate for a Matter Dominated Universe}

Now let us use the result of the last section to (as an example)
calculate the local Hubble expansion rate for a matter-dominated Universe. 
In this case, the scale factor $a$ and the scalar metric perturbation
$\Phi$ have the following dependence on the conformal time $\eta$ :
\begin{eqnarray}
a(\eta) \, &=& \, a_{m}\eta^{2}/2  \label{smfactor}  \\
\Phi(\eta,x) \, &=& \, C_{1}(x)+C_{2}(x)\eta^{-5} \label{smpert} \, ,
\end{eqnarray}
where $a_m$ is a constant, and $C_1$ and $C_2$ are time-independent.
The second equation is valid in the long wavelength (super-Hubble-scale)
limit and has been explicitly derived in
\cite{RA}. Now using equations (\ref{expform}), (\ref{smfactor}) and
(\ref{smpert}) we obtain
$\Theta $ in terms of conformal time:
\begin{eqnarray}
\Theta \, &=& \, \frac{12}{a_{m}}\eta^{-3}(1-C_{1}(x)+\frac{3}{2}C_{1}^{2})
\nonumber \\
&-& \frac{2}{3}\frac{1}{a_{m}}\eta^{-1}((\partial_{i}C_{1}(x))^{2} 
+ \partial_{i}^{2}C_{1}(x)) \nonumber \\ 
&+& \frac{1}{a_{m}}\eta^{-6}
(\frac{1}{3}(\partial_{i}C_{1}(x))(\partial_{i}C_{2}(x))
+ \partial_{i}^{2}C_{2}(x)) \\
&+& \frac{1}{a_{m}}\eta^{-8}(3C_{2}(x)+36C_{1}(x)C_{2}(x)) \nonumber \\
&+& \frac{1}{a_{m}}\eta^{-11}
(\partial_{i}C_{2}(x))^{2}+\frac{1}{a_{m}}\eta^{-13}(C_{2}(x))^{2} \nonumber
\end{eqnarray}
Some of the terms are decreasing very fast as a function of time and thus we
can ignore them. If we just keep the terms with the powers $-3$
and $-1$ of $\eta$ then we get:
\begin{eqnarray} \label{matterfin}
\Theta \, &=& \, \frac{12}{a_{m}}\eta^{-3}(1-C_{1}(x)+\frac{3}{2}C_{1}^{2}) \\
&-& \frac{2}{3}\frac{1}{a_{m}}\eta^{-1}[(\partial_{i}C_{1}(x))^{2}
+\partial_{i}^{2}C_{1}(x)] \nonumber
\end{eqnarray}

If we take the average of $\Theta$ on a constant $\eta$ hypersurface,
only the terms quadratic in the fluctuation variables survive.
Thus, considering large values of $\eta$ and focusing on the second
order terms, it appears from (\ref{matterfin}) that infrared modes
give a positive contribution to $\Theta$ and thus lead to a
speeding-up of the expansion, whereas 
ultraviolet modes enter with a negative sign and thus yield a slowing 
effect, the latter
becoming more significant (relative to the unperturbed expansion rate)
for larger values of $\eta$.

However, before drawing definite physical conclusions from our analysis,
we must take into account that the background time
$\eta$ is not an observable quantity. To obtain results for
back-reaction of any real physical significance  we have to find
an observable variable like proper time and evaluate the expansion rate
in terms of this variable, so that we can discuss its evolution from an
observer's point of view. 

If we use equation (\ref{metric}) for the metric, 
equation (\ref{smpert}) for $\Phi$, and
equation (\ref{smfactor}) for the scale factor, 
we can find the expression for the proper time $\tau$ in
terms of conformal time. Since
\begin{equation}
d\tau^{2}=a(\eta)^{2}(1+2\Phi)d\eta^{2} \, ,
\end{equation}
a simple integration yields
\begin{eqnarray}
\tau \, &=& \, \frac{a_{m}}{6}(1+C_{1}-\frac{1}{2}C_{1}^{2})\eta^{3} \\
&-& \frac{a_{m}}{4}(C_{2}-C_{1}C_{2})\eta^{-2}
+ \frac{a_{m}}{28}C_{2}^{2}\eta^{-7} \nonumber
\end{eqnarray}
In the approximation of large values of $\eta$ we can ignore the
second and the third term of this equation and thus obtain:
\begin{equation}
\eta^{-3} \, = \, \frac{a_{m}}{6}(1+C_{1}-\frac{1}{2}C_{1}^{2} )\tau^{-1} \, .
\end{equation}
Now we can use this relation and substitute into the first term of
equation (\ref{matterfin}). We see the effects of the dominant infrared 
terms on the local expansion rate cancel exactly up to second order in
perturbations, when evaluating $\Theta$ at a fixed proper time:
\begin{equation}
\Theta_{IR}=2\tau^{-1} \, .
\end{equation}
This implies that at least in the approximation of keeping only
the leading infrared terms, there is no local gravitational
back-reaction of cosmological fluctuations on the local Hubble
expansion rate in this matter-dominated universe.

\section{Expansion Rate in Terms of Scalar Field as an Observable }

We now move on to the example more relevant to the work of 
\cite{bran2,bran3,aw0}, namely a Universe dominated by a single
real scalar field $\varphi$,which is a toy model for inflationary cosmology.
During inflation, fluctuations which are generated on sub-Hubble
scales early on during the inflationary phase are red-shifted to
scales much larger than the Hubble radius. Thus, in this context
it is of great interest to consider the back-reaction of infrared modes.

In the following, we will generalize the previous analysis to
be applicable to matter consisting of a single scalar field. In this case
one can treat the scalar field as a
perfect fluid and derive the velocity four-vector field. To do this, we
need to write the energy-momentum tensor $T_{\mu\nu}$ of the scalar field
(right hand side of the following equation) in the form of an
energy-momentum tensor for a perfect fluid (left hand side of the following
equation):
\begin{equation}
(\rho+P)u_{\mu}u_{\nu} - P g_{\mu\nu} \, 
= \, \partial_{\mu}\varphi\partial_{\nu}\varphi-{\cal L} \, .
g_{\mu\nu}
\end{equation}
At the level of the background fields, the two expressions
are identical if we take  $P = {\cal L}$ and 
$u_{\mu}=A\partial_{\mu}\varphi$ with 
\begin{equation}
A \, = \,  \bigl(\partial^{\nu}\varphi\partial_{\nu}\varphi\bigr)^{-1/2}
\, .
\end{equation}

Now that we have shown that the energy-momentum tensor of a
scalar field can be written in the form of that of a perfect fluid,
we can use the expression (\ref{expform}) which gives $\Theta$ in terms of the
metric fluctuations to evaluate the local effect of gravitational
back-reaction of cosmological fluctuations. Let us first for convenience
rewrite $\Theta$ of (\ref{expform}) in
terms of the physical time $t$:
\begin{equation} \label{expform2}
\Theta \, = \, 3\frac{\dot{a}}{a}(1-\Phi+\frac{3}{2}\Phi^{2})
-3\dot{\Phi}+\frac{(\partial_{i}\dot{\Phi})(\partial_{i}\Phi)+\partial^{2}_{i}\dot{\Phi}}{a\ddot{a}-\dot{a}^{2}}
\end{equation}

Theoretically, the scalar field $\varphi$ is an observable. In fact,
in a system with a single matter field $\varphi$, it is this field
which must be used as a clock. Hence, to obtain results with physical
meaning, we must evaluate $\Theta$ on a surface of constant $\varphi$
and not constant $t$.

As discussed above, in the context of inflationary cosmology
it is important to study the effects of infrared modes  up to
second order. We will also assume that we are in the slow rolling 
regime of inflation. In our case (as well as for a more general case),
the prescription is to calculate $t$, $\Phi$, $a$ and
$\frac{\partial}{\partial t}$ in terms of $\varphi$, and to insert
the results into the general expression (\ref{expform2}) for $\Theta$. 
The relation between $t$ and $\varphi$ can be derived starting from:
\begin{equation}
\varphi(t)=\varphi_{0}(t)+\delta \varphi_{1}(t) \, ,
\end{equation}
which can be written as:
\begin{equation}
t \, = \, \varphi_{0}^{-1}[\varphi-\delta \varphi_{1}(t)] \, .  
\end{equation}
Thus, the equation
\begin{eqnarray}
t \, &=& \, \varphi_{0}^{-1}(\varphi)-\frac{\partial\varphi_{0}^{-1}(\varphi)}{\partial\varphi}\delta\varphi_{1}(\varphi_{0}^{-1}(\varphi)) \nonumber \\
&+&(\frac{\partial\varphi_{0}^{-1}(\varphi)}{\partial\varphi})^{2}\frac{\partial\delta\varphi_{1}(\varphi_{0}^{-1}(\varphi))}{\partial
t}\delta\varphi_{1}(\varphi_{0}^{-1}(\varphi)) \\
&+& \frac{1}{2}\frac{\partial^{2}\varphi_{0}^{-1}(\varphi)}{\partial
\varphi^{2}}\delta \varphi_{1}^{2}(\varphi_{0}^{-1}(\varphi)) \nonumber
\end{eqnarray}
relates $t$ and $\varphi$. 

We now wish to express the value of $\Theta$ 
in terms of $\varphi$ (note that we are considering
the local value of $\Theta$ in this analysis, and there is no
need to perform a spatial averaging).
In order to relate the metric, its fluctuations and the scale factor 
to the scalar field and its fluctuations   
we need to take an explicit form for the potential and make use of
the Einstein constraint equations. The
$G^{0i}$ and $G^{00}$ equations relate $\Phi$ to $\varphi$ (at the level
of the first order fluctuations) and the Hubble parameter of the background
to the background scalar field $\varphi_{0}$ (at the level
of the unperturbed Friedmann equations).
In our simple system we do not need to go through
these calculations explicitly (as long as we are
interested only in the leading infrared terms) since by dropping the 
gradient terms from the
$G^{0i}$ and $G^{00}$ equations, it can be shown that \cite{niayesh}
\begin{equation}
\frac{H}{\sqrt{1+2\Phi}} \, = \, {1 \over {\sqrt{3}}}\sqrt{V(\varphi)} \, .
\end{equation}
If we expand the right hand side in terms of $\Phi$ we get the
result
\begin{equation}
H(1-\Phi+\frac{3}{2}\Phi^{2}) \, 
= \, {1 \over {\sqrt{3}}}\sqrt{V(\varphi)} \, ,
\end{equation}
and substituting this result into equation (\ref{expform2}) and neglecting
terms containing $\dot{\Phi}$ and $\partial_{i}\Phi$ (which are
sub-dominant compared to other terms in the infrared and slow roll limit)
leads to the final result for the local expansion rate:
\begin{equation} \label{final}
\Theta \, = \, \sqrt{3} \sqrt{V(\varphi)}
\end{equation}
which as a function of $\varphi$ is the same as the relation
for an unperturbed background. 

Thus, again we do not see any 
back-reaction of cosmological perturbations on the local expansion rate
in this approximation. In retrospect it is easy to understand this
result, since \cite{niayesh} by neglecting gradient terms we have a
Friedmann Universe whose expansion rate satisfies equation (\ref{final}).

\section{Conclusions}

In this paper, we have studied the back-reaction effects on a local
observable which measures the local expansion rate of the Universe.
The observable gives the rate at which neighboring comoving observers
separate and coincides with the usual definition of expansion in
the context of the fluid approach to cosmology. In order to obtain
a physical quantity, we evaluated the observable at a fixed value of
the scalar field.

We evaluated our observable, the {\it local physical expansion
rate}, in a simple toy model of chaotic inflation consisting of a
single scalar matter field coupled to gravity. We found that the
leading infrared terms, the terms which dominate the effects
discussed in \cite{bran2} and \cite{aw0}, cancel if we evaluate
the observable at a fixed value of the scalar field, whereas they
do not vanish if we evaluate them at a fixed value of the
background time. The former result is a physical result since it
corresponds to a physical observable evaluated at a space-time
point specified by a physical prescription, whereas the latter
result (obtained by evaluating at a fixed background time) does
not have a diffeomorphism-invariant meaning. Our analysis thus
confirms the concern of \cite{unruh} that the results obtained in
\cite{bran2} and \cite{aw0} are not invariant under second order
gauge transformations. Our results confirm the conclusions of
\cite{aw3} reached by means of a different method of analysis
applied to a different physical observable.

Our result does {\bf not} imply that there is no back-reaction of
the infrared modes of cosmological perturbations. There is no
reason to expect that the next to
leading infrared terms in our result will cancel (they do not
cancel in the analysis of \cite{aw3}). One of the
advantages of our technique is that they can be evaluated without
too much trouble. This is left to a future publication. So, even
in single field models of inflation there might be some
non-vanishing back-reaction of infrared modes.

We expect that back-reaction of infrared modes will be much more
important in two field models of inflation. Let us assume that the
matter sector of the theory contains both an inflation field
$\varphi$ and a regular matter field $\chi$ (with
nonvanishing and time-dependent spatial average) which, for example,
could represent the cosmic microwave background. In this case, it
is no longer true that long wavelength fluctuations have no
physical effects on local observables. If the measurement point
is (in an unambiguous physical way) determined by a fixed value
of the field $\chi$, then the local expansion rate may
sensitively depend on the amplitude of the long wavelength
fluctuations in $\varphi$. Thus, the leading infrared terms may
{\bf not} cancel when evaluated according to the abovementioned
physical prescription in the same way that they do not cancel in
the analysis of this paper when the observable is evaluated at a
fixed value of the background time.

There is a close analogy with the analysis of the parametric
amplification of super-Hubble-scale cosmological fluctuations
during inflationary reheating. From the point of view of the
background space-time coordinates, it appears \cite{Kaiser} that
the parametric amplification of matter fluctuations on
super-Hubble scales in an unperturbed cosmological background (see
e.g. \cite{tb,kls} for a discussion of parametric resonance during
reheating) would imply the parametric amplification of the
cosmological fluctuations on these scales. However, it can be
shown that in single field models physical observables measuring
the amplitude of cosmological fluctuations do not feel any
resonance \cite{fb1,parry,lin,niayesh}. In contrast, in two field models
of inflation there is \cite{bv,fb2} parametric amplification of
super-Hubble-scale cosmological fluctuations. In this case, there is a
fluctuation mode corresponding to entropy fluctuations which
cannot locally be gauged away. This mode is (in certain theories)
parametrically amplified during reheating, and in turn drives the
parametric resonance of the super-Hubble scale curvature
fluctuations.

Our techniques allow us to calculate the back-reaction of cosmological
fluctuations in two field models in a very similar way to what is
presented here. Results will be presented in a followup publication
\cite{Ghazal2}.

\medskip
\centerline{Acknowledgments}
\medskip

We would like to thank Bill Unruh, Alan Guth and Andrei Linde for
discussions which stimulated this research. We are grateful to
Raul Abramo, Niayesh Afshordi, Yasusada Nambu and Richard Woodard 
for many useful
insights during the course of this research. This research is
supported in part (at Brown) by the US Department of Energy under
Contract DE-FG0291ER40688, Task A. One of us (GG) is grateful to the Department
of Astrophysical Sciences of Princeton University for hospitality
during the course of this work. The other of us (RB) wishes to
thank the CERN Theory Division and the Institut d'Astrophysique
de Paris for hospitality and financial support.

\end{document}